\documentclass[a4paper,10pt]{article}
\usepackage[english]{babel}
\usepackage{amsmath}
\usepackage{amsthm}
\usepackage{amssymb,units,mathtools}
\usepackage[T1]{fontenc}
\usepackage{commath}
\usepackage{vwcol}
\usepackage[a4paper]{geometry}

\usepackage[utf8]{inputenc}
\usepackage{graphicx}
\usepackage{mathrsfs}
\usepackage{xfrac}
\usepackage{multicol}
\usepackage{caption}
\usepackage{subcaption}
\usepackage{multirow}
\usepackage{graphicx}
\usepackage{mathrsfs}
\usepackage{xfrac}
\usepackage{fancyhdr}
\usepackage{comment}
\usepackage{multicol}
\usepackage{multirow}
\usepackage{float}
\usepackage[overload]{empheq}
\usepackage{epstopdf}
\usepackage{epsfig}
\usepackage[nobysame]{amsrefs}
\usepackage{xfrac}
\usepackage{float}
\usepackage{abstract}
\usepackage[symbol]{footmisc}

\newcommand{\R}{\mathbb{R}}

\usepackage{pst-node}
\usepackage{tikz-cd} 
\usepackage{xcolor}
\usepackage{color}
\usepackage{titlesec}
\usepackage{chapterbib}
\titleformat{\chapter}[display]
{\normalfont\huge\bfseries}{\chaptertitlename\ \thechapter}{20pt}{\Huge}
\titlespacing*{\chapter}{0pt}{-10pt}{20pt}

\usepackage[colorlinks = true]{hyperref}

\hypersetup{
	linkcolor=black,
	citecolor = black,
	urlcolor=black,
	filecolor=black
}

\title{Application of Herglotz's Variational Principle to Electromagnetic Systems with Dissipation}

\author{\sffamily 
\sc $^a$Jordi Gaset
\thanks{jordi.gaset@unir.net\,({\it ORCID}:\,0000-0001-8796-3149).} ,
$^b$Adrià Marín-Salvador\thanks{
   adria.marin@st-hughs.ox.ac.uk (ORCID: 0000-0001-8054-1576).}
\\[1ex]
\normalsize\itshape\sffamily 
$^a$Escuela Superior de Ingeniería y Tecnología, Universidad Internacional de La Rioja, Spain.
\\[1ex]
\normalsize\itshape\sffamily 
$^b$ Mathematical Institute, University of Oxford, United Kingdom.
}
\date{\today}

\begin{document}
\maketitle

\begin{abstract}

This work applies the contact formalism of classical mechanics and classical field theory, introduced by
Herglotz and later developed in the context of contact geometry, to describe electromagnetic systems with
dissipation. In particular, we study an electron in a non-perfect conductor and a variation of the cyclotron
radiation. In order to apply the contact formalism to a system governed by the Lorentz force, it is necessary
to generalize the classical electromagnetic gauge and add a new term in the Lagrangian.
We also apply the k-contact theory for classical fields to model the behaviour of electromagnetic fields
themselves under external damping. In particular, we show how the theory describes the evolution of
electromagnetic fields in media under some circumstances.  The corresponding Poynting theorem is derived. We discuss its applicability to the Lorentz
dipole model and to a highly resistive dielectric.

\end{abstract}

\medskip


\noindent\textbf{MSC\,2020 codes:}
37K58, 37L05, 53D10, 35Q53

\tableofcontents

\section{Introduction}

Around 1930, G. Herglotz introduced a variational method to model mechanical systems with dissipation \cites{Herglotz_1930, Herglotz_1985}. Herglotz allowed the Lagrangian of the system to depend on the action itself, and obtained a set of generalized Euler-Lagrange equations that happened to be useful when considering some dissipative phenomena. The framework of such description was later discovered to be contact geometry \cites{Geiges_2008, Etnyre_2001}.

Recently, there has been a renewed interest in contact geometry due to its success in modelling several systems. Some of these applications include thermodynamics \cites{Mrugala1991,Grillo2020,Bravetti2019, Simoes2020, Simoes2020b}, statistical mechanics \cite{Bravetti2016}, geometric optics \cite{Arnold1990}, hydrodynamics \cite{Etnyre_2000}, circuit theory \cite{Goto2016} and  control theory \cites{Vassiliou_2010,Leon2020}.

In parallel, contact geometry has been expanded with new structures and frameworks. The Lagrangian formulation and its symmetries have been studied \cites{Gaset2020,Gaset_2020newcon,de_Le_n_2019singularandprecontact, de_Le_n_2020infinitessimal,de_Le_n_2019contactsystems}, and several generalizations have been developed: the higher-order case \cite{Leon2021HO}, the Hamilton-Jacobi theory \cite{DeLeon2021HJ} and the unified formalism \cite{DeLeon2020Unified}.

Recently, a higher-dimensional analogue of contact geometry has been proposed, called $k$-contact geometry, which can be applied to model field theories with dissipation \cites{Gaset_2020contactframework, Gaset_2021}. This description generalizes previous attempts to define an action principle for action-dependant Lagrangian field densities \cite{Lazo_2018}.

The present work applies the contact formalism of classical mechanics and classical field theory to electromagnetism. This has, to the best of our knowledge, not been done explicitly in literature, apart from a small example in \cite{Lazo_2018}. Electromagnetism provides an interesting context in which to apply the theory. On the one hand, the study of a particle under a Lorentz force allows for the use of the contact formalism of classical mechanics. On the other, one can focus on the study of the evolution of the electromagnetic vector fields themselves applying the $k$-contact geometry theory. In addition, one can think of multiple examples in which a particle moves under the influence of a Lorentz force in the presence of external dissipation \cite{Razavy2016}*{Ch. 6}. It is also known how electromagnetic fields can be damped when in a medium \cite{Haus1989}*{Sec. 11.5}.

It should be clear that contact geometry does not provide a description of all dissipative phenomena, but rather simply produces a set of equations of motion from a Lagrangian function that is action-dependant. Thus, not all dissipative systems can be explained via this theory and its use is not constrained to such models. This work explores to what extent dissipation in electromagnetic systems can be modelled using contact formalism.

During this work, some objects will be referred to as \textit{dissipative} although they actually represent a break in energy conservation, which might be an increase or a decrease of the energy. It will we specified when this is the case and when we actually refer to energy loss.

Driven by previous successful applications of contact geometry in dissipation \cites{Gaset_2020newcon,Gaset_2021,Gaset_2020contactframework, Lazo_2018, de_Le_n_2019singularandprecontact}, we have decided to produce the equations of motion from a Lagrangian function which is the classical symplectic Lagrangian plus a linear term in the action. We allow the linear term to be tuned a posteriori in order to fit the experimental results in every particular example.

The currently developed contact techniques for classical mechanics and classical field theory only apply to autonomous Lagrangians. That is, in the classical mechanical case, the Lagrangians cannot depend explicitly on time, and in the $k$-contact framework, the Lagrangian densities cannot depend explicitly on the components of the spacetime. This is the reason for which during this work it will be asked that the objects appearing in the Lagrangian do not have explicit terms in such components. However, using variational principles, one can show that the produced equations of motion are still valid in the non-autonomous case. Future research is needed in order to develop the mathematical tools to describe such Lagrangians using the contact formalism.

This work is structured as follows. Section \ref{geometry} presents a mathematical introduction to Lagrangian contact systems and Lagrangian k-contact systems. In Section \ref{Lorentz}, we apply the contact formalism of classical mechanics to study particles under a Lorentz force. When producing the equations of motion from the classical Lagrangian of a Lorentz force, one finds that they are not gauge invariant. This can be solved by slightly changing the Lagrangian and introducing a new characterization of the electromagnetic gauge, which reduces to the classical one for non-dissipative systems. The theory is applied to two particular examples, an electron in a non-perfect conductor and a particle in a magnetic field with damping. Finally, in section \ref{ElectromagneticFields} we apply the $k$-contact theory to electromagnetic fields. Apart of the dissipative term, we consider a particular metric to model lineal materials. In some cases, the classical Lagrangian density in vacuum can model electromagnetic fields in media when adding a linear term in the action. We derive sufficient and necessary conditions that a system must satisfy so that the theory applies. Finally, the applicability of the produced equations of motion is discussed in two real-life examples. 

\section{Contact and k-contact Lagrangian systems}
\label{geometry}

Lagrangian contact systems are an special case of contact systems, which we will introduce briefly. The reader can find a more detailed exposition in \cites{Gaset_2020newcon,de_Le_n_2020infinitessimal,de_Le_n_2019singularandprecontact}.

Consider the manifold $TQ\times\mathbb{R}$, where $Q$ is an $n$ dimensional manifold which represents the configuration space of the system, with local coordinates $(q^i,v^i,s)$. The canonical endomorphism $\mathcal{S}$ and the Liouville vector field $\Delta$ of $TQ$ extend to $TQ\times\mathbb{R}$ in the usual way. Their local expressions are
\begin{align*}
    \mathcal{S}&= \frac{\partial}{\partial v^i}\otimes dq^i\,,
    \\
    \Delta&=v^i \frac{\partial}{\partial v^i}\,.
\end{align*}
A Lagrangian is a function $\mathcal{L}:TQ\times\mathbb{R}\rightarrow\mathbb{R}$. The associated contact form is $\eta_\mathcal{L}=ds-\mathcal{S}^*(d\mathcal{L})$, with local expression
\begin{align*}
    \eta_\mathcal{L}=ds-\frac{\partial\mathcal{L}}{\partial v^i}dq^i\,.
\end{align*}
The Reeb vector field of $\eta_\mathcal{L}$ is
\begin{equation*}
    \mathcal{R}_\mathcal{L}=\frac{\partial}{\partial s}-W^{ij}\frac{\partial^2 \mathcal{L}}{\partial v^j\partial s}\frac{\partial}{\partial v^i}\,
\end{equation*}
where $(W^{ij})$ is the inverse of the Hessian of $\mathcal{L}$. The associated Lagrangian energy is $E_\mathcal{L}=\Delta(\mathcal{L})-\mathcal{L}$. Its local expression is
\begin{align*}
    E_\mathcal{L}=v^i \frac{\partial \mathcal{L}}{\partial v^i}-\mathcal{L}\,.
\end{align*}
Then, we have the contact system $(\eta_\mathcal{L}, E_\mathcal{L})$. Its solutions are integral curves of a vector field  $X\in\mathfrak{X}(TQ\times\mathbb{R})$, which is a SODE, and satisfies the generalized Euler-Lagrange equations
\begin{align*}
    \iota_X d\eta_\mathcal{L}&= d E_\mathcal{L}-\left(L_{\mathcal{R}_\mathcal{L}}E_\mathcal{L}\right)\eta_\mathcal{L}\,,
    \\
    \iota_X\eta_\mathcal{L}&=-E_\mathcal{L}\,.
\end{align*}
For a holonomic curve $\sigma(t)=(q^i(t), v^i(t),s(t))$, these equations take the local expression 
\begin{align}\nonumber
    \dot{v}^j\frac{\partial^2 \mathcal{L}}{\partial v^i \partial v^j}+v^j\frac{\partial^2 \mathcal{L}}{\partial v^i \partial q^j}+\dot{s}\frac{\partial^2 \mathcal{L}}{\partial v^i \partial s}-\frac{\partial \mathcal{L}}{\partial q ^i}&=\frac{\partial \mathcal{L}}{\partial v^i }\frac{\partial \mathcal{L}}{\partial s}\,,
    \\\label{sisaction}
    \dot{s}&=\mathcal{L}\,.
\end{align}

Hence,  $s$ can be interpreted as the action of the system. We are therefore modelling systems in which the Lagrangian depends on the action itself.

The generalized Euler-Lagrange equations can also be obtained via a variational method, as seen in \cite{de_Le_n_2019singularandprecontact}*{Section 5}, which was introduced by G. Herglotz in \cite{Herglotz_1930}. Let $\mathcal{L}\in \mathscr{C}^\infty(TQ\times\R)$ be a Lagrangian function and consider two points $x,y\in Q$. Consider $[0,1]\subseteq\R$ and let us denote by $\mathscr{S}$ the space of all smooth curves
$
\alpha:[0,1]\rightarrow Q
$
such that $\alpha(0) = x$ and $\alpha(1) = y$. 

Let us denote by $\mathscr{C}^\infty([0,1]\rightarrow X)$ the set of all smooth mappings from $[0,1]$ to a manifold $X$. One can define the functional
\[
\begin{array}{cccc}
	\mathcal{Z}:&\mathscr{C}^\infty([0,1]\rightarrow Q)&\rightarrow&\mathscr{C}^\infty([0,1]\rightarrow \R)\\
	&\xi&\mapsto&\mathcal{Z}(\xi),
\end{array}
\]
where $\mathcal{Z}(\xi)$ is the unique solution to
\begin{align}[left = \empheqbiglbrace]
	\label{ODEvariational}
	&\frac{d\mathcal{Z}(\xi)(t)}{dt} = \mathcal{L}(\xi(t), \dot{\xi}(t), \mathcal{Z}(\xi)(t))\\
	&\mathcal{Z}(\xi)(0) = 0.\nonumber
\end{align}

After the previous discussion on the problem and particularly Equation \eqref{sisaction}, it is clear why one would define such an operator $\mathcal{Z}$: given a curve $\xi$, its image $\mathcal{Z}(\xi)$ is the action of the system associated to the path $\xi$. Hence, the physically realisable path between $x$ and $y$ is the curve $\xi\in\mathscr{S}$ that minimizes
\begin{equation*}
	\mathcal{Z}(\xi)(1),
\end{equation*}	
the action at the endpoint, see \cite{de_Le_n_2019singularandprecontact}*{Thm. 2}. Note that, from Equation \eqref{ODEvariational}, we find
\begin{equation}
	\label{VariationalIntegral}
\mathcal{Z}(\xi)(1) = \int_0^1\mathcal{L}(\xi(t), \dot{\xi}(t), \mathcal{Z}(\xi)(t))dt.
\end{equation}

Let $\eta(t) = \big(\eta_1(t),\eta_2(t), \eta_3(t)\big)\in\mathscr{C}^\infty([0,1]\rightarrow Q)$ such that $\eta(0) = \eta(1) = 0$ and consider $\mathcal{Z}(\xi+\varepsilon\eta)(1)$. The variational problem \eqref{VariationalIntegral} implies that
\begin{equation}
	\label{VariationalCondition}
0 = \frac{d\mathcal{Z}(\xi+\varepsilon\eta)(1)}{d\varepsilon}\Big|_{\varepsilon = 0} = \int_0^1\Big(\frac{\partial\mathcal{L}}{\partial q_i}\eta_i+\frac{\partial\mathcal{L}}{\partial v_i}\dot{\eta}_i\Big)dt + \int_0^1\frac{\partial\mathcal{L}}{\partial s}\frac{d\mathcal{Z}(\xi+\varepsilon\eta)(t)}{d\varepsilon}\Big|_{\varepsilon = 0} dt.
\end{equation}

Let us now define $\zeta(r) = \frac{d\mathcal{Z}(\xi+\varepsilon\eta)(r)}{d\varepsilon}|_{\varepsilon = 0}$ and $A(t) = \frac{\partial\mathcal{L}}{\partial q_i}\eta_i+\frac{\partial\mathcal{L}}{\partial v_i}\dot{\eta}_i$. Note that, for $r\in(0,1]$,
\begin{equation}
\label{ImposeVariational}
\zeta(r) = \int_0^rA(t)dt+\int_0^r\frac{\partial\mathcal{L}}{\partial s}\zeta(t)dt,
\end{equation}
which implies
\[
\frac{d\zeta}{dr}(r) =  A(r)+\frac{\partial \mathcal{L}}{\partial s}\zeta(r).
\]

Hence, solving the ODE with the initial condition $\zeta(0) = 0$,
\begin{equation*}
	\zeta(r) = \exp\Big(\int_0^r\frac{\partial\mathcal{L}}{\partial s}(\theta)d\theta\Big)\cdot\int_0^r\exp\Big(-\int_0^t \frac{\partial\mathcal{L}}{\partial s}(\theta)d\theta\Big)A(t)dt.
\end{equation*}

Let us impose that $\zeta(1) = 0$, which is equivalent to imposing that the action reaches a relative extremum. Since the first factor cannot vanish, this reads
\begin{equation*}
	0 = \int_0^1\exp\Big(-\int_0^t \frac{\partial\mathcal{L}}{\partial s}(\theta)d\theta\Big)A(t)dt = \int_0^1\exp\Big(-\int_0^t \frac{\partial\mathcal{L}}{\partial s}(\theta)d\theta\Big)\Big(\frac{\partial\mathcal{L}}{\partial q_i}\eta_i+\frac{\partial\mathcal{L}}{\partial v_i}\dot{\eta}_i\Big)dt,
\end{equation*}
and integrating by parts the second term
\begin{align*}
	 0 &=   \int_0^1\exp\Big(-\int_0^t \frac{\partial\mathcal{L}}{\partial s}(\theta)d\theta\Big)\frac{\partial\mathcal{L}}{\partial q_i}\eta_idt +  \int_0^1\exp\Big(-\int_0^t \frac{\partial\mathcal{L}}{\partial s}(\theta)d\theta\Big)\frac{\partial\mathcal{L}}{\partial v_i}\dot{\eta}_idt = \\ &=  \int_0^1\exp\Big(-\int_0^t \frac{\partial\mathcal{L}}{\partial s}(\theta)d\theta\Big)\frac{\partial\mathcal{L}}{\partial q_i}\eta_idt -\int_0^1\exp\Big(-\int_0^t \frac{\partial\mathcal{L}}{\partial s}(\theta)d\theta\Big)\Big[\frac{d}{dt}\Big(\frac{\partial\mathcal{L}}{\partial v_i}\Big)-\frac{\partial\mathcal{L}}{\partial s}\frac{\partial\mathcal{L}}{\partial v_i}\Big]\eta_idt = \\ & = \int_0^1\exp\Big(-\int_0^t \frac{\partial\mathcal{L}}{\partial s}(\theta)d\theta\Big)\Big[\frac{\partial\mathcal{L}}{\partial q_i}-\frac{d}{dt}\Big(\frac{\partial\mathcal{L}}{\partial v_i}\Big)+\frac{\partial\mathcal{L}}{\partial s}\frac{\partial\mathcal{L}}{\partial v_i}\Big]\eta_idt,
\end{align*}
which, by the fundamental lemma of  calculus of variations \cite{J.RgenJost2007}*{Lemm. 1.1.1}, implies the generalized Euler-Lagrange equations
\begin{equation*}
	\frac{\partial\mathcal{L}}{\partial q_i}-\frac{d}{dt}\Big(\frac{\partial\mathcal{L}}{\partial v_i}\Big)+\frac{\partial\mathcal{L}}{\partial s}\frac{\partial\mathcal{L}}{\partial v_i} = 0.
\end{equation*}

\subsection{k-contact Lagrangian systems}
\label{kcontact}

The k-contact structure was introduced in \cite{Gaset_2020contactframework} in order to generalize contact mechanics to field theories. The Lagrangian formalism, which we use in Section \ref{ElectromagneticFields}, was developed in \cite{Gaset_2021}. In this section the principle elements of this formalism are stated. Moreover, we will give a variational formulation of the $k$-contact Euler-Lagrange equations.

The Lagrangian k-contact formalism of a system with an $k$-dimensional configuration space $Q$ over an $k$-dimensional space-time takes place in $\oplus^kTQ\times\mathbb{R}^k$. The bundle $\oplus^kTQ$ over $Q$ is the Whitney sum of $k$ copies of the tangent bundle, each one representing the derivative of the coordinate over the different directions of space-time. Moreover, $k$ dissipative variables are considered. Natural coordinates on $\oplus^kTQ\times\mathbb{R}^k$ are $(q^i,q^i_\mu,s^\mu)$, where $1\leq i\leq n$ and $1\leq \mu \leq k$.

A Lagrangian is a function $\mathcal{L}:\oplus^kTQ\times\mathbb{R}^k\rightarrow\mathbb{R}$. On this work we only consider a particular class of Lagrangians, those with linear dependence on $s^\mu$ with constant coefficients. 

The k-contact structure is formed by the $k$ $1$-forms
$$
\theta^\mu_\mathcal{L}=ds^\mu-\frac{\partial \mathcal{L}}{\partial q^i_\mu}dq^i\,.
$$
The Reeb vector fields are a set of $k$ vector fields such that 
\begin{equation}\label{eq:reeb}
  \iota_{\mathcal{R}^\alpha}\theta^\beta=\delta^\beta_\alpha\,;\quad \iota_{\mathcal{R}^\alpha}d\theta^\beta=0 \,.  
\end{equation}
For the particular class of Lagrangians we will consider in this work, they can be chosen to be $\mathcal{R}_\mu=\frac{\partial}{\partial s^\mu}$. Finally, the Lagrangian energy is given by
$$
E_\mathcal{L}=\frac{\partial \mathcal{L}}{\partial q^i_\mu}q^i_\mu-\mathcal{L}.
$$

The solutions are holonomic functions $\sigma:\mathbb{R}^k\rightarrow \oplus^kTQ\times\mathbb{R}^k$, which are integrable sections of a $k$-dimensional distribution. This distribution can be described by $k$ vector fields $(X_\mu)$ and a section $\sigma$ is integral of $(X_\mu)$ if 
$$
T\sigma \circ \frac{\partial }{\partial x^\mu}=X_\mu\circ\sigma\,.
$$
 The k-contact Euler-Lagrange equations are
 \begin{align}\label{eq:kcontact1}
    \iota_{X_\mu} d\theta^\mu_\mathcal{L}&= d E_\mathcal{L}-\left(L_{\mathcal{R}_\mu}E_\mathcal{L}\right)\theta^\mu_\mathcal{L}\,,
    \\\nonumber
    \iota_{X_\mu}\theta^\mu_\mathcal{L}&=-E_\mathcal{L}\,.
\end{align}
For a holonomic function $\sigma=(q^i(x^\mu),q^i_\mu(x^\mu),s^\mu(x^\mu)$, these equations take the local expression
 \begin{align*}
    \frac{\partial q^j_{\nu}}{\partial x^\mu}\frac{\partial^2 \mathcal{L}}{\partial q^i_\mu \partial q^j_\nu}+q^j_\mu\frac{\partial^2 \mathcal{L}}{\partial q^i_\mu \partial q^j}+ \frac{\partial s^\nu}{\partial x^\mu}\frac{\partial^2 \mathcal{L}}{\partial q^i_\mu \partial s^\nu}-\frac{\partial \mathcal{L}}{\partial q ^i}&=\frac{\partial \mathcal{L}}{\partial q^i_\mu }\frac{\partial \mathcal{L}}{\partial s^\mu}\,,
    \\
    \frac{\partial s^\mu}{\partial x^\mu}&=\mathcal{L}\,.
\end{align*}
The Lagrangian $k$-contact formalism presented in \cite{Gaset_2021} is developed for regular Lagrangians. Unfortunately, the Lagrangian used in section \ref{ElectromagneticFields} is singular. Nevertheless, one can try to follow all the steps described above, but there is one problem: Equations \eqref{eq:reeb} do not have a unique solution for singular Lagrangians. To circumvent this issue, in section \ref{ElectromagneticFields} we will use an idea introduced in \cite{de_Le_n_2019singularandprecontact} for the mechanical case: to proof that Equations \eqref{eq:kcontact1} are independent of the solution of \eqref{eq:reeb} chosen.

Just like in the classical contact formulation, it is possible to derive the $k$-contact Euler-Lagrange equations for fields as a result of a variational principle. Let $\mathcal{L}(q^i, q^i_\mu, s^\mu)$ be a Lagrangian function on $\oplus^k TQ\times \R^k$. A field on $Q$ is a smooth map
\[
\begin{array}{cccc}
	\Psi:&\Omega\subset\R^k&\rightarrow& Q\times \mathbb{R}^k\\
	& x^\nu &\rightarrow& (\Psi_\mu(x^\nu),\Psi_{s^\mu}(x^\nu)\equiv s^\mu(x^\nu) )
\end{array}
\]
from an open subset $\Omega$, and the action related to such field is defined as
\[
S(\Psi) = \int_\Omega\mathcal{L}(\Psi_\mu, \partial_\nu\Psi_\mu, s^{\mu})d^4x.
\]
Then, the equations of motion can be obtained by minimizing the action $S(\Psi)$ with respect to $\Psi$ under the constraint
\[
\partial_\mu s^{\mu} = \mathcal{L}(\Psi_\mu, \partial_\nu\Psi_\mu, s^{\mu}).
\]
By the Lagrange multiplier Theorem for Banach spaces \cite{Luenberger1968}*{Thm. 9.3.1}, extremizing $S$ with the above constraint is equivalent to extremizing the following function with respect to $\Psi$,
\begin{align*}
f(\Psi_\mu, \partial_\nu\Psi_\mu, s^\mu,\partial_\nu s^\mu, \lambda ) &= \int_\Omega\Big[\mathcal{L}(\Psi_\mu, \partial_\nu\Psi_\mu, s^\mu)-\lambda(x^\alpha)\big(\partial_\mu s^\mu-\mathcal{L}(\Psi_\mu, \partial_\nu\Psi_\mu, s^\mu)\big)\Big]d^4x \\&= \int_\Omega\Big[\partial_\mu s^\mu-\lambda(x^\alpha)\big(\partial_\mu s^\mu-\mathcal{L}(\Psi_\mu, \partial_\nu\Psi_\mu, s^\mu)\big)\Big]d^4x,
\end{align*}
where $\lambda:\Omega\rightarrow\R$ is a smooth function which we call Lagrange multiplier function. The Euler-Lagrange equations for $f$ read
\begin{align*}[left = \empheqbiglbrace]
	\partial_\nu\Big(\lambda(x^\alpha)\frac{\partial\mathcal{L}}{\partial(\partial_\nu\Psi_\mu)}\Big)&=\lambda(x^\alpha)\frac{\partial\mathcal{L}}{\partial\Psi_\mu}\\
	-\partial_\mu\lambda(x^\alpha) &= \lambda(x^\alpha)\frac{\partial\mathcal{L}}{\partial s^\mu}\\
	0&=\partial_\mu s^\mu-\mathcal{L}(\Psi_\mu, \partial_\nu\Psi_\mu, s^\mu),
\end{align*}
where the last equation comes from imposing that the Euler-Lagrange equations are also satisfied for $\lambda$.

Expanding the first equation,
\begin{align*}
	\lambda(x^\alpha)\frac{\partial\mathcal{L}}{\partial\Psi_\mu} &= \frac{\partial\mathcal{L}}{\partial(\partial_\nu\Psi_\mu)}\partial_\nu\lambda(x^\alpha)+\lambda(x^\alpha)\partial_\nu\Big(\frac{\partial\mathcal{L}}{\partial(\partial_\nu\Psi_\mu)}\Big)\\&=-\lambda(x^\alpha)\frac{\partial\mathcal{L}}{\partial(\partial_\nu\Psi_\mu)}\frac{\partial\mathcal{L}}{\partial s^\nu}+\lambda(x^\alpha)\partial_\nu\Big(\frac{\partial\mathcal{L}}{\partial(\partial_\nu\Psi_\mu)}\Big),
\end{align*}
which, dividing by $\lambda(x^\alpha)$, implies
\[
\partial_\nu\Big(\frac{\partial\mathcal{L}}{\partial(\partial_\nu\Psi_\mu)}\Big)-\frac{\partial\mathcal{L}}{\partial\Psi_\mu} = \frac{\partial\mathcal{L}}{\partial(\partial_\nu\Psi_\mu)}\frac{\partial\mathcal{L}}{\partial s^\nu},
\]
the $k$-contact Euler-Lagrange equations for fields.

\section{Study of Particles under a Lorentz Force}
\label{Lorentz}

\subsection{Symplectic Formulation}
\label{SymplecticEM}
\subsubsection{The Equations of Motion}
\label{SymplecticEoM}
Let $Q\subseteq\R^3$ be an open subset of $\R^3$ on which a particle of mass $m$ and charge $k$ moves under the influence of the electromagnetic force $$\textbf{F} = k(\textbf{E}+\textbf{v}\times\textbf{B})$$ induced by an electric field $\textbf{E}$ and a magnetic field $\textbf{B}$ which are time independent. Let $\textbf{q} = (q_1,q_2,q_3)$ be coordinates defined on $Q$. Assume, in addition, that the electromagnetic potentials
\[
\begin{array}{cccc}
	\phi: &Q&\rightarrow&\R\\
	&\textbf{q}&\mapsto &\phi(\textbf{q})
\end{array}
\hspace{2cm}
\begin{array}{cccc}
	\textbf{A}: &Q&\rightarrow&\R^3\\
	&\textbf{q}&\mapsto &\textbf{A}(\textbf{q})
\end{array}
\]
that define $\textbf{E}$ and $\textbf{B}$ are also time independent. Recall that $\textbf{A}$ and $\phi$ are such that 
\[
\textbf{E} = -\nabla\phi\hspace*{1cm}\text{and}\hspace*{1cm}\textbf{B} = \nabla\times\textbf{A}.
\]
The Lagrangian associated with the Lorentz force is
\[
\mathcal{L} = \frac{m}{2}\textbf{v}\cdot\textbf{v}+k\textbf{A}\cdot\textbf{v}-k\phi.
\]
The Lagrangian energy function and the Lagrangian symplectic form read
\[
E_\mathcal{L} = \frac{\partial \mathcal{L}}{\partial \textbf{v}}\cdot\textbf{v}-\mathcal{L} = \big(m\textbf{v}+k\textbf{A}\big)\cdot\textbf{v}-\mathcal{L} = \frac{m}{2}\textbf{v}\cdot\textbf{v}+k\phi
\]
and
\[
\omega_\mathcal{L} = -\sum\limits_{i = 1}^3d\Big(\frac{\partial\mathcal{L}}{\partial v_i}\Big)\wedge dq_i = -\sum\limits_{i = 1}^3d(mv_i+kA_i)\wedge dq_i = m\sum\limits_{i = 1}^3 dq_i\wedge dv_i+k\sum\limits_{i = 1}^3\Big(\sum\limits_{j\neq i}\frac{\partial A_i}{\partial q_j} dq_i\wedge dq_j\Big).
\]

The dynamics of the Lagrangian dynamical system $(TQ, \omega_\mathcal{L}, E_\mathcal{L})$ are encoded in the vector field $X_\mathcal{L}\in\mathfrak{X}(TQ)$ solution to
\begin{equation}
	\label{EquationFieldEoMLorentz}
	\iota_{X_\mathcal{L}}\omega_\mathcal{L} = dE_\mathcal{L}.
\end{equation} 
One can compute
\[
dE_\mathcal{L} = k\frac{\partial\phi}{\partial \textbf{q}}\cdot d\textbf{q}+m\textbf{v}\cdot d\textbf{v}.
\]
If we write
\[
X_\mathcal{L} = \sum\limits_{i = 1}^3Q_i\frac{\partial}{\partial q_i}+\sum\limits_{i = 1}^3V_i\frac{\partial}{\partial v_i},
\]
then Equation \eqref{EquationFieldEoMLorentz} reads
\begin{align}
	m\sum\limits_{i = 1}^3Q_idv_i+\sum\limits_{i = 1}^3\Bigg(-mV_i+k\sum\limits_{j\neq i}\Big(\frac{\partial A_j}{\partial q_i}-\frac{\partial A_i}{\partial q_j}\Big)Q_j\Bigg)dq_i = k\sum\limits_{i = 1}^3\frac{\partial\phi}{\partial q_i}dq_i+m\sum\limits_{i = 1}^3v_idv_i.
\end{align}

Since all the differentials are linearly independent, the previous equation implies
\begin{align}[left = \empheqbiglbrace]
	Q_i &= v_i\\
	mV_i &= k\sum\limits_{j\neq i}\Big(\frac{\partial A_j}{\partial q_i}-\frac{\partial A_i}{\partial q_j}\Big)Q_j-k\frac{\partial \phi}{\partial q_i}.
\end{align}

Hence, the time evolution of the particle in configuration space is given by $(\textbf{q}(t), \textbf{v}(t))$ with $\dot{\textbf{q}}(t) = \textbf{v}(t)$, and
\begin{equation}
\label{EoMLorentzSymplecti }
m\dot{\textbf{v}}(t) = k\textbf{v}(t)\times(\nabla\times\textbf{A})-k\nabla\phi,
\end{equation}
which are, of course, the equations of motion obtained by using the Euler-Lagrange equations. Note that we find $m\dot{\textbf{v}} = k\textbf{E}+k\textbf{v}\times\textbf{B}$, which are the equations of motion obtained by applying Newton's second law to the Lorentz force.
\subsubsection{Classical Gauge}
\label{ClassicalGauge}
The choice of vector and scalar potentials at the beginning of Section \ref{SymplecticEoM} is not unique. Indeed, let $f\in\mathscr{C}^\infty(Q\times\R)$ be a smooth function on $Q\times\R$, where the $\R$ coordinate depicts time, and let
\[
\phi' = \phi-\frac{\partial f}{\partial t}\hspace{1cm}\text{and}\hspace*{1cm}\textbf{A}' = \textbf{A}+\nabla f.
\]
Assume that the primed potentials are also time independent, which is equivalent to imposing
\[
\frac{\partial^2 f}{\partial t^2} = \frac{\partial^2f}{\partial t\partial q_i} = 0.
\]

Then, $$\nabla\phi' = \nabla\phi-\nabla\frac{\partial f}{\partial t} = \nabla\phi-\sum\limits_{i = 1}^3\frac{\partial^2 f}{\partial q_i\partial t} = \nabla\phi$$ and $$\nabla\times\textbf{A}' = \nabla\times\textbf{A}+\nabla\times(\nabla f) = \nabla\times\textbf{A},$$ and hence the induced electric and magnetic fields remain unchanged. This freedom in choosing the vector and scalar potentials is known as gauge freedom, and the choice of a particular pair $(\phi,\textbf{A})$ is called a gauge fixing or a choice of gauge. It is believed that the gauge is not measurable since, as deduced in Section \ref{SymplecticEoM}, the equations of motion only depend on the observable fields $\textbf{E}$ and $\textbf{B}$.

Although the equations of motion remain unchanged when the gauge changes, the Lagrangian does not. Indeed, let us compute
\begin{align*}
\mathcal{L}' = \frac{m}{2}\textbf{v}\cdot\textbf{v}+k\textbf{A}'\cdot\textbf{v}-k\phi' = \mathcal{L}+k(\nabla f)\cdot\textbf{v}+k\frac{\partial f}{\partial t} = \mathcal{L}+\frac{d(kf)}{dt}.
\end{align*}

It is known that the addition of a full time derivative to the Lagrangian does not change the extrema of the action and hence it does not change the equations of motion. However, it is important to note that the Lagrangian and the action themselves do depend on the choice of gauge.

\subsection{Contact Formulation}
\label{ContactEM}
\subsubsection{A First Attempt at the Equations of Motion}
Consider again a particle of mass $m$ and charge $k$ moving in an open subset $Q\subseteq\R^3$ under the influence of a Lorentz force $\textbf{F} = k(\textbf{E}+\textbf{v}\times\textbf{B})$. Assume $\textbf{E}$ and $\textbf{B}$ are time independent and let $(\phi,\textbf{A})$ be a choice of gauge for the electric and magnetic fields such that the potentials are also time independent. Driven by previous successful applications of contact formalism to mechanical systems with dissipation \cites{Gaset_2020newcon, de_Le_n_2019singularandprecontact}, we propose the following contact Lagrangian on $TQ\times\R$:
\[
\mathcal{L} = \frac{m}{2}\textbf{v}\cdot\textbf{v}+k\textbf{A}\cdot\textbf{v}-k\phi-\gamma s
\]
for some $\gamma\in\R$.

The Lagrangian energy associated to this system is
\[
E_\mathcal{L} = \frac{\partial\mathcal{L}}{\partial\textbf{v}}\cdot\textbf{v}-\mathcal{L} = (m\textbf{v}+k\textbf{A})\cdot\textbf{v}-\mathcal{L} = \frac{m}{2}\textbf{v}\cdot\textbf{v}+k\phi+\gamma s,
\]
and the canonical contact form equals
\[
\eta_\mathcal{L} = ds-\frac{\partial\mathcal{L}}{\partial \textbf{v}}\cdot d\textbf{q} = ds-m\textbf{v}\cdot d\textbf{q}-k\textbf{A}\cdot d\textbf{q}.
\]

Since all the second derivatives $\frac{\partial^2\mathcal{L}}{\partial s\partial v_i}$ vanish, the Reeb vector field is simply $\mathcal{R} = \frac{\partial}{\partial s}$. It is clear that the directional derivative of $E_\mathcal{L}$ in the direction of $\frac{\partial}{\partial s}$ is the constant $\gamma$. Thus, the generalized Euler-Lagrange equations simplify to
\begin{align}
	\label{ContactVectorFieldEM}
\iota_{X_\mathcal{L}}d\eta_\mathcal{L} &= dE_\mathcal{L}-\gamma\eta_\mathcal{L}\\
\label{ContactVectorFieldEM2}
\iota_{X_\mathcal{L}}\eta_\mathcal{L} &= -E_\mathcal{L}
\end{align}
for an unknown vector field $X_\mathcal{L}\in\mathfrak{X}(TQ\times \R)$. Let us compute the differentials
\[
dE_\mathcal{L} = k\nabla\phi\cdot d\textbf{q}+m\textbf{v}\cdot d\textbf{v}+\gamma ds
\]
and
\[
d\eta_\mathcal{L} = m\sum\limits_{i = 1}^3 dq_i\wedge dv_i+k\sum\limits_{i,j = 1}^3\frac{\partial A_j}{\partial q_i}dq_j\wedge dq_i.
\]

If we write the unknown vector field $X_\mathcal{L}$ in coordinates as
\[
X_\mathcal{L} = \sum\limits_{i = 1}^3 Q_i\frac{\partial}{\partial q_i}+\sum\limits_{i = 1}^3V_i\frac{\partial}{\partial v_i} + S\frac{\partial}{\partial s},
\]
then Equation \eqref{ContactVectorFieldEM2} reads
\[
S-\frac{\partial\mathcal{L}}{\partial \textbf{v}}\cdot \textbf{Q} = -\frac{\partial\mathcal{L}}{\partial \textbf{v}}\cdot\textbf{v}+\mathcal{L}
\]
and if we impose that the system is holonomic, that is $Q_i = \dot{q}_i = v_i$, then the solution curve $(\textbf{q}(t), \textbf{v}(t), s(t))$ will satisfy $$\dot{s} = \mathcal{L},$$
and the Lagrangian is interpreted to depend on the action $\int\mathcal{L}dt$ itself.

If we now let $\dot{v}_i = V_i$, Equation \eqref{ContactVectorFieldEM} implies 
\[
-m\dot{\textbf{v}}\cdot d\textbf{q}+k\sum\limits_{i = 1}^3\sum\limits_{j\neq i}\Big(\frac{\partial A_j}{\partial q_i}-\frac{\partial A_i}{\partial q_j}\Big)Q_jdq_i = k\nabla\phi\cdot d\textbf{q}+\gamma (m\textbf{v}+k \textbf{A})\cdot d\textbf{q},
\]
and given the linear independence of the 1-forms $dq_i$, 
\begin{equation}
	\label{EoMFirstContcat}
	m\dot{\textbf{v}}(t) = -k(\nabla\times\textbf{A})\times\textbf{v}(t)-k\nabla\phi-\gamma(m\textbf{v}(t)+k\textbf{A}).
\end{equation}
Note that we find the same two terms as in Equation \eqref{EoMLorentzSymplecti }, together with a dissipative term in velocities $-\gamma m\textbf{v}(t)$ and an interaction term between the dissipation and the vector potential, $-\gamma k\textbf{A}$.

The obtained equations of motion are not invariant under the classical gauge defined in Section \ref{ClassicalGauge}. Indeed, if we make a change of gauge
\[
\phi' = \phi-\frac{\partial f}{\partial t}\hspace{1cm}\text{and}\hspace*{1cm}\textbf{A}' = \textbf{A}+\nabla f
\]
between time independent potentials, the first two terms remain invariant, but the interaction term does not, and hence then the equations of motion become
\[
m\dot{\textbf{v}}(t) = -k(\nabla\times\textbf{A})\times\textbf{v}(t)-k\nabla\phi-\gamma(m\textbf{v}(t)+k\textbf{A})-\gamma k\nabla f.
\]

Note that a change of gauge with a smooth function $f$ introduces a full time derivative in the Lagrangian, as discussed in Section \ref{ClassicalGauge}. Hence, we observe that, in the contact framework, adding a total time derivative does not, in general, preserve the equations of motion. The Lorentz force provides an example of how producing equivalent Lagrangians in the contact framework differs from the symplectic case.

Since we want the equations of motion to be gauge invariant, and the gauge to be non-observable, we will introduce a new characterization of the gauge and propose a new contact Lagrangian in the next section.
\subsubsection{Generalized Gauge}
\label{GeneralizedGauge}
In this section we introduce a new characterization of the classical gauge that generalizes the one defined in Section \ref{ClassicalGauge}. Instead of a pair $(\phi,\textbf{A})$, a choice of gauge will now be a triplet
\[
(\phi, \textbf{A}, f),
\]
where $(\phi,\textbf{A})$ is a classical gauge and $f\in\mathscr{C}^\infty(Q)$ is a smooth function on $Q$. As discussed in previous sections, only time independent scalar and vector potentials are considered. Note that, since $f$ is a function on $Q$, it is time independent. 

We will define two choices of gauge $(\phi,\textbf{A}, f),\ (\phi',\textbf{A}',f')$ to be related by the gauge if there exists a smooth function $g\in\mathscr{C}^\infty(Q\times\R)$ such that
\begin{align*}[left = \empheqbiglbrace]
\phi'	 &= \phi-\frac{\partial g}{\partial t}\\
\textbf{A}' &= \textbf{A}+\nabla g\\
f' &= f-g.
\end{align*}

For both gauges to be time independent, it will be necessary and sufficient that $\frac{\partial g}{\partial t} =  0$. Note that this is equivalent to $g\in\mathscr{C}^\infty(Q)$ and that it implies that the scalar potential remains unchanged. Note also that the relation defined on the choices of gauge is an equivalence relation, and hence it produces equivalence classes of choices of gauge. Let $[(\phi,\textbf{A},f)]$ denote the equivalence class of the choice of gauge $(\phi, \textbf{A}, f)$.

We claim that, in each class $[(\phi,\textbf{A}, f)]$, there exists a unique choice of gauge of the type $(\phi',\textbf{A}',0)$. Indeed,
\[
[(\phi,\textbf{A}, f)] = [(\phi-\frac{\partial f}{\partial t},\textbf{A}+\nabla f, f-f)] = [(\phi-\frac{\partial f}{\partial t},\textbf{A}+\nabla f, 0)],
\]
and if $[(\phi,\textbf{A}, 0)] = [(\phi',\textbf{A}', 0)]$ the function $g\in\mathscr{C}^\infty(Q)$ that relates them must satisfy $0 = 0-g$, and hence must be $g = 0$. Then, $\phi = \phi'$ and $\textbf{A} = \textbf{A}'$,

We propose, for a choice of gauge $(\phi, \textbf{A}, f)$, the contact Lagrangian
\begin{equation}
	\label{SecondContactLagrangian}
	\mathcal{L} = \frac{m}{2}\textbf{v}\cdot \textbf{v}+k\textbf{A}\cdot \textbf{v}+k\frac{\partial f}{\partial\textbf{q}}\cdot\textbf{v}-k\phi-\gamma s,
\end{equation}
where we explicitly make use of the function $f\in\mathscr{C}^\infty(Q)$ of the gauge.

\subsubsection{The Equations of Motion Revisited}
\label{ContactEM2}
Assume a particle of mass $m$ and charge $k$ is moving in an open subset $Q\subseteq\R^3$ under the influence of an electromagnetic force. Let $(\phi, \textbf{A}, f)$ be a choice of gauge as defined in Section \ref{GeneralizedGauge}, and let the contact Lagrangian of the system be the one defined in Equation \eqref{SecondContactLagrangian},
\begin{equation}\label{eq:lagrevisited}
	\mathcal{L} = \frac{m}{2}\textbf{v}\cdot \textbf{v}+k\textbf{A}\cdot \textbf{v}+k\frac{\partial f}{\partial \textbf{q}}\cdot\textbf{v}-k\phi-\gamma s.
\end{equation}
The Lagrangian energy density associated with this Lagrangian is
\[
E_\mathcal{L} = \frac{\partial\mathcal{L}}{\partial \textbf{v}}\cdot\textbf{v}-\mathcal{L} = \frac{m}{2}\textbf{v}\cdot\textbf{v}+k\phi+\gamma s,
\]
and the contact 1-form is
\[
\eta_\mathcal{L} = ds-\frac{\partial\mathcal{L}}{\partial \textbf{v}}\cdot d\textbf{q} = ds-m\textbf{v}\cdot d\textbf{q}-k\textbf{A}\cdot d\textbf{q}-k\frac{\partial f}{\partial \textbf{q}}\cdot d\textbf{q}.
\]

Just like in Section \ref{ContactEM}, the Reeb vector field is $\mathcal{R}_\mathcal{L} =  \frac{\partial}{\partial s}$, and the directional derivative of the energy with respect to the Reeb vector field is $\mathscr{L}_{\mathcal{R}_\mathcal{L}}E_\mathcal{L} = \frac{\partial E_\mathcal{L}}{\partial s}=  \gamma$. Hence, the generalized Euler-Lagrange equations read
\begin{align}
		\label{ContactVectorFieldSecondEM}
	\iota_{X_\mathcal{L}}d\eta_\mathcal{L} &= dE_\mathcal{L}-\gamma\eta_\mathcal{L}\\
	\label{ContactVectorFieldSecondEM2}
	\iota_{X_\mathcal{L}}\eta_\mathcal{L} &= -E\mathcal{L}.
\end{align}
A straightforward computation gives
\[
dE_\mathcal{L} = m\textbf{v}\cdot d\textbf{v}+k\nabla\phi\cdot d\textbf{q}+\gamma ds
\]
and
\[
d\eta_\mathcal{L} = m\sum\limits_{i = 1}^3 dq_i\wedge dv_i+k\sum\limits_{i,j = 1}^3\Big(\frac{\partial A_j}{\partial q_i}\Big)dq_j\wedge dq_i.
\]

Since the only difference with respect to Equations \eqref{ContactVectorFieldEM} and \eqref{ContactVectorFieldEM2} is the extra term $- k\nabla f\cdot d\textbf{q}$ in the contact form, the equations of motion are
\[
\dot{s} = \mathcal{L}
\]
and
\begin{equation}
	\label{EoMSecondContact}
m\dot{\textbf{v}}(t) = -k(\nabla\times\textbf{A})\times\textbf{v}(t)-k\nabla\phi-\gamma\big(m\textbf{v}(t)+k\textbf{A}\big)-\gamma k\nabla f,
\end{equation}
once we have imposed that the system is holonomic, i.e. $\dot{\textbf{q}} = \textbf{v}$.

We claim that Equation \eqref{EoMSecondContact} is now independent of the choice of gauge. Indeed, if we take $g\in\mathscr{C}^\infty(Q)$ and let
\begin{align*}[left = \empheqbiglbrace]
	\phi'	 &= \phi\\
	\textbf{A}' &= \textbf{A}+\nabla g\\
	f' &= f-g,
\end{align*}
Equation \eqref{EoMSecondContact} becomes
\begin{align*}
m\dot{\textbf{v}}(t) &= -k(\nabla\times\textbf{A})\times\textbf{v}(t)-k\nabla\phi-\gamma\big(m\textbf{v}(t)+k\textbf{A}\big)-\gamma k\nabla g-\gamma k\nabla f+\gamma k\nabla g = \\ & =-k(\nabla\times\textbf{A})\times\textbf{v}(t)-k\nabla\phi-\gamma\big(m\textbf{v}(t)+k\textbf{A}\big)-\gamma k\nabla f,
\end{align*}
and thus, it remains unchanged. Actually, note that the Lagrangian \eqref{SecondContactLagrangian} is itself invariant under a change of gauge, unlike in the symplectic case.

In this new proposed framework, the observable fields are
\begin{align}[left = \empheqbiglbrace]
	\label{Observables}
\textbf{E} &= -\nabla\phi\nonumber\\
\textbf{R}& = \textbf{A}+\nabla f \\
\textbf{B} &= \nabla\times\textbf{A} = \nabla \times \textbf{R}
\nonumber
\end{align}
for a choice of gauge $(\phi, \textbf{A}, f)$, and the equations of motion are
\[
m\dot{\textbf{v}} = k\textbf{v}(t)\times\textbf{B}+k\textbf{E}-\gamma m\textbf{v}(t)-\gamma k\textbf{R},
\]
where all three fields are invariant under the gauge. Note that knowing $\textbf{R}$ allows us to know $\textbf{A}$ up to a gradient, which is exactly the same freedom for $\textbf{A}$ when knowing $\textbf{B}$. In addition, the generalized moment of a particle under the proposed Lagrangian is
\[
\frac{\partial\mathcal{L}}{\partial \textbf{v}} = m\textbf{v}+k\textbf{A}+k\nabla f = m\textbf{v}+k\textbf{R},
\]
which is also an observable, whilst in the symplectic case the generalized moment is not an observable.

\subsubsection{Equivalent Lagrangians}

In order to have a more complete understanding of the gauge freedom of the Lagrangian, let us analyze its equivalent Lagrangians.

Two Lagrangians are equivalent if they lead to the same solutions. In symplectic mechanics, two Lagrangians that differ by a total derivative are equivalent. In contact mechanics, one can construct equivalent Lagrangians by considering transformations for the $s$ variable, which can be thought of as constructing new actions with the same critical points. One can generalize the symplectic result to the contact setting by considering transformations of the form $s\rightarrow \zeta(q^i,s)=s+h(q^i)$.  

For a Lagrangian $\mathcal{L}$ and a transformation $\zeta$, the corresponding equivalent Lagrangian is given by:
$$
\bar{\mathcal{L}}(x,q,v,\zeta)=\mathcal{L}(x,q,v,s)+\nabla h\cdot \mathbf{v}\,.
$$
Notice that the action variable for the new Lagrangian is $\zeta$. Considering the Lagrangian \eqref{eq:lagrevisited} we are interested in, for any function $h(q^i)$ we have an equivalent Lagrangian
$$
\bar{\mathcal{L}}(x,q,v,\zeta)= \frac{m}{2}\textbf{v}\cdot \textbf{v}+k\textbf{A}\cdot \textbf{v}+k\nabla f\cdot\textbf{v}-k\phi-\gamma (\zeta-h)+\nabla h\cdot \mathbf{v}\,.
$$
Setting $h=kf$ we have a particularly interesting equivalent Lagrangian:
\begin{equation}\label{eq:lagrevisited2}
\bar{\mathcal{L}}(x,q,v,\zeta)= \frac{m}{2}\textbf{v}\cdot \textbf{v}+k\textbf{A}\cdot \textbf{v}-k\phi-\gamma (\zeta+kf)\,,
\end{equation}
which is another gauge invariant realization for a particle moving with dissipation under a Lorentz force given by $(\phi,\mathbf{A},f)$. Indeed, performing a gauge transformation given by a function $g$, we obtain
$$
\bar{\bar{\mathcal{L}}}(x,q,v,\zeta)= \frac{m}{2}\textbf{v}\cdot \textbf{v}+k\textbf{A}\cdot \textbf{v}+k\nabla g\cdot \textbf{v}-k\phi-\gamma (\zeta-kg+kf)\,,
$$
which is an equivalent Lagrangian to \eqref{eq:lagrevisited2} given by the transformation $s\rightarrow s+kg$.

When performing a gauge transformation on the Lagrangian \eqref{eq:lagrevisited2} we obtain an equivalent Lagrangian, as in the symplectic case. On the other hand, a gauge transformation leaves \eqref{eq:lagrevisited} invariant, with no need to invoke equivalence theorems. Notice that this is also the case when we recover the symplectic case setting $\gamma=0$. The triplet description of the electromagnetic gauge $(\phi,\mathbf{A},f)$, together with the term $k\textbf{A}\cdot \textbf{v}+k\nabla f\cdot\textbf{v}-k\phi$ in the Lagrangian gives us a variational description of the Lorentz force where the Lagrangian is invariant under gauge transformations (in both the classical and the contact settings).

\subsubsection{Example. Electron in a non-Perfect Conductor}
We shall now apply the formalism discussed in Section \ref{ContactEM2} to describe the motion of an electron in a non-perfect conductor. 

Consider a sufficiently large but finite cylindrical non-perfect conductor of length $L$ and cross-section $A$. Let $\sigma$ denote the conductivity of the material. Assume a voltage $\Delta V$ is applied between the ends of the conductor, which is known to generate an electric field of constant magnitude
\[
E = -\frac{\Delta V}{L}
\]
in the longitudinal direction of the conductor. We take this direction to correspond to be the $x$ coordinate. Assume no magnetic fields intervene in the problem. Hence, one can take the vector potential to be $\textbf{A} =0 $ and the gauge to be
\[
(\phi, \textbf{A}, f)  = \Big(\frac{\Delta V}{L}x,0,0\Big)
\]
to describe the problem. Indeed, the observables for this choice as described in Equation \eqref{Observables} are 
\[
\textbf{E} = -\frac{\Delta V}{L}\textbf{e}_x\hspace*{1cm}\text{and}\hspace*{1cm}\textbf{B} = \textbf{R}= 0,
\]
where $\textbf{e}_x$  denotes the unit vector in the $x$ direction. The equations of motion \eqref{EoMSecondContact} read
\begin{equation}
\dot{v}(t) = -\mu\frac{\Delta V}{L}-\gamma v(t)
\end{equation}
in the $x$ direction, where $\mu:=\frac{k_e}{m_e}<0$, for $k_e$ and $m_e$ the charge and mass of the electron respectively. If we assume the electron starts at rest at $x  = 0$, the previous ODE can be solved for the velocity to find
\begin{equation}
	\label{VelocityElectron}
v(t) = \frac{\mu\Delta V}{\gamma L}\Big(e^{-\gamma t}-1\Big),
\end{equation}
and Equation \eqref{VelocityElectron} can be integrated to obtain
\[
x(t) = -\frac{\mu\Delta V}{\gamma^2L}(\gamma t+e^{-\gamma t}-1).
\]
Note that the electron reaches a limit velocity
\begin{equation}
	\label{LimitVelocity}
v_{lim} = \lim\limits_{t\rightarrow\infty}v(t) = -\frac{\mu\Delta V}{\gamma  L}>0.
\end{equation}
Given that the conductor is ohmic, the limit velocity (or drift velocity) of an electron in the conductor can be shown to be \cite{EdwardM.Purcell2019}*{p. 187}
\[
v_{lim} = -\frac{m_{mol}\sigma\Delta V}{\rho k_e n L},
\]
where $m_{mol}$ is the molecular mass of the conductor, $\rho$ is the density of the conductor and $n$ the number of free electrons per molecule of conductor. Hence, one can impose the limit velocity in \eqref{LimitVelocity} is the drift velocity of the electron to find
\[
\gamma  = \frac{\mu k_e\rho n}{m_{mol}\sigma}.
\]

If we take copper as the material the conductor is made of, one finds that $\gamma\approx 4\cdot 10^{13}\unit{s^{-1}}$ and hence the limit velocity is achieved at the order of $t\approx 10^{-13}\unit{s}$. This means that the electron has travelled approximately $3\cdot10^{-16}\unit{m}$. Thus, the assumption that $L$ is large enough so that the limit velocity is achieved is physically realizable.

The energy of the electron in the described electric field is known to be,
\[
E = \frac{m_e}{2}v^2+k_e\frac{\Delta V}{L}x
\]
and hence it dissipates at a rate
\begin{align*}
\frac{dE}{dt} &= m_e v(t)\dot{v}(t)+k_e\frac{\Delta V}{L}v(t)\\ &= -\frac{k_e\mu(\Delta V)^2}{\gamma L^2}\Big(e^{-\gamma t}-1\Big)e^{-\gamma t}+\frac{k_e\mu(\Delta V)^2}{\gamma L^2}\Big(e^{-\gamma t}-1\Big) = \\
& = -\frac{k_e\mu(\Delta V)^2}{\gamma L^2}\Big(e^{-\gamma t}-1\Big)^2\,,
\end{align*}
which, as $t\rightarrow\infty$, when the drift velocity is achieved, becomes
\begin{equation}
\label{EnergyDisspationElectron}
\frac{dE}{dt} = -\frac{k_e\mu(\Delta V)^2}{\gamma L^2} = -\frac{m_{mol}\sigma (\Delta V)^2}{n\rho L^2}.
\end{equation}

Assume there is no interaction between the electrons in the conductor, which implies that all of them dissipate energy at the same rate when they reach the drift velocity. The number of electrons in the conductor is given in the variables of the problem by
\[
n_e = n \frac{M_T}{m_{mol}} = \frac{n\rho AL}{ m_{mol}},
\]
where $M_T$ denotes the total mass of the conductor. Hence, the total energy dissipation rate within the conductor when all electrons have achieved the drift velocity is
\[
\frac{dE_T}{dt} = n_e\frac{dE}{dt} = -\frac{A\sigma}{L}(\Delta V)^2,
\]
which is precisely Joule's heating law.
\subsubsection{Example. Particle in a Magnetic Field}
Consider a particle of charge $k$ and mass $m$, in a magnetic field $\textbf{B}  = (0,0,B)$. A choice of vector potential for $\textbf{B}$ is
\[
\textbf{R} = \textbf{A} = \frac{B}{2}(-y,x,0),
\]
taking $f = 0$, and if we assume no electric fields are present, the equations of motion \eqref{EoMSecondContact} read
\begin{align*}[left = \empheqbiglbrace]
	\dot{v}_x &= \mu Bv_y-\gamma v_x+\frac{\gamma\mu}{2}By\\
	\dot{v}_y & = -\mu B v_x -\gamma v_y-\frac{\gamma\mu}{2}Bx,
\end{align*}
where $\mu: = \frac{k}{m}$. This system of equations is not easy to discuss for arbitrary values of $\gamma$. However, if we define the energy of the particle as its kinetic energy, we see that
\begin{align*}
	\frac{d E}{dt} &=\frac{m}{2}\frac{d}{dt}(v_x^2+v_y^2) =  m(v_x\dot{v}_x+v_y\dot{v}_y) = m\Big(v_x( \mu Bv_y-\gamma v_x+\frac{\gamma\mu}{2}By)+v_y( -\mu B v_x -\gamma v_y-\frac{\gamma\mu}{2}Bx)\Big) \\ &= -\gamma m(v_x^2+v_y^2)+\frac{m\gamma\mu B}{2}(v_xy-xv_y) = -\gamma m v^2-\frac{\gamma\mu }{2}\textbf{B}\cdot \textbf{L}.
\end{align*}

 It is known that a charged particle spinning in a magnetic field which is perpendicular to its velocity experiences a dissipation of its energy due to the emission of radiation \cite{Monreal_2016}. This effect is known as cyclotron radiation. The frequency of a particle of mass $m$ emitting cyclotron radiation in the classical limit is 
\[
f = \frac{kB}{2\pi m},
\]
and the energy dissipated due to the cyclotron radiation satisfies \cite{Longair_1994}*{Eq. 18.8}
\[
\frac{dE}{dt} = -\frac{\sigma_t B^2 v^2}{c\mu_0},
\]
where $\sigma_t = \frac{8\pi}{3}\Big(\frac{k^2}{4\pi\epsilon_0 m c^2}\Big)^2$ is the Thomson total cross-section.

In order to argue about the nature of the motion of the particle, let us assume that the cross terms $\frac{\gamma\mu}{2}Bx$ and $\frac{\gamma\mu}{2}By$ can be neglected. Then, the system reads
\begin{align*}[left = \empheqbiglbrace]
	\dot{v}_x &= \mu Bv_y-\gamma v_x\\
	\dot{v}_y & =- \mu B v_x -\gamma v_y,
\end{align*}
and can be solved by
\begin{align*}[left = \empheqbiglbrace]
	v_x(t) &= e^{-\gamma t}\big(v_{x0}\cos(\mu B t)+v_{y0}\sin(\mu B t)\big)\\
	v_y(t) & = e^{-\gamma t}\big(v_{y0}\cos(\mu B t)-v_{x0}\sin(\mu B t)\big),
\end{align*}
from which we deduce that the particle describes a decreasing spiral in the plane if $\gamma>0$. Note that the frequency of the motion is exactly the frequency of a particle emitting cyclotron radiation.

Let the energy of the particle be
 \[
 E(t) =m\frac{v^2(t)}{2} =  m\frac{v_x^2+v_y^2}{2} = m\frac{e^{-2\gamma t}}{2}v_0^2,
 \]
 where $v_0^2 = v_{x_0}^2+v_{y_0}^2$ is the square of its initial velocity. Then, the energy is dissipated at a rate
 \[
 \frac{dE}{dt} = -m\gamma e^{-2\gamma t}v_0^2 = -m\gamma v^2,
 \]
 and we only obtain the term in the cyclotron dissipation. Let us impose that the dissipated energy of the model is precisely the dissipation in the cyclotron radiation. Then,
 \begin{equation}
 \label{gammacyclotron}
 \gamma = \frac{\sigma_t B^2}{ c\mu_0 m}>0.
 \end{equation}
 
 If we take the particle to be an electron, then $\mu \approx -1.76\cdot 10^{11}\unit{C/kg}$. If we let $B= 1\unit{T}$, then $\gamma \approx 0.1938\unit{s}^{-1}$. Let us return now to the general case, considering the cross-terms. Let us assume we can fix $\gamma$ as in Equation \eqref{gammacyclotron} in order to estimate the solutions of the system. Assume that $0<\gamma \ll|\mu B|$. If we let $\eta = x+iy$, then
 \[
 \ddot{\eta} = -\dot{\eta}\big(i\mu B+\gamma\big)-\frac{i\gamma\mu B}{2},
 \] 
 and the eigenvalues of the characteristic polynomial are
 \begin{align*}[left = \empheqbiglbrace]
 	 \frac{-\gamma-i\mu B}{2}+\frac{i}{2}\sqrt{\mu^2B^2-\gamma^2}&\approx -\frac{\gamma}{2}-\frac{i\gamma^2}{4\mu B} \\
 	 \frac{-\gamma-i\mu B}{2}-\frac{i}{2}\sqrt{\mu^2B^2-\gamma^2}&\approx-\frac{\gamma}{2}-i\mu B+\frac{i\gamma^2}{4\mu B} .
 \end{align*}

The frequency of the motion is altered in the order $\mathcal{O}(\frac{\gamma^2}{\mu B})$. In addition, if $0<\gamma\ll |\mu B|$,  the particle will describe a slightly perturbed decreasing spiral motion for small times. The term
\[
-\frac{\gamma\mu}{2}\textbf{B}\cdot\textbf{L}
\]
is always positive, but it can be seen that, when $0<\gamma\ll|\mu B|$, it is smaller in norm than the dissipative term due to the cyclotron radiation. Hence, the particle does indeed lose energy, but at a lower rate than in the cyclotron. We are modelling the small-time behaviour of a particle inside a magnetic field that dissipates energy due to the emission of a cyclotron radiation that has been altered by the interaction between the external magnetic field and the angular momentum of the particle itself.

\begin{figure}[H]
    \centering
    \begin{subfigure}[t]{0.45\textwidth}
        \centering
       \includegraphics[height=5.5cm]{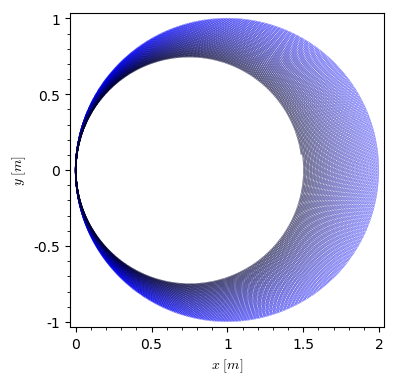}
    \end{subfigure}
    \begin{subfigure}[t]{0.45\textwidth}
        \centering
       \includegraphics[height=5.5cm]{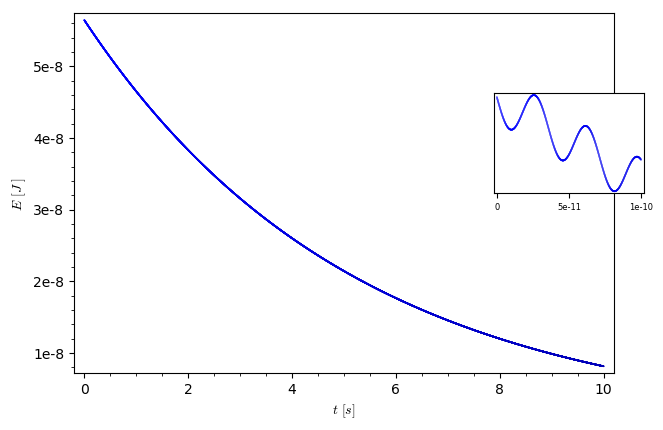}
    \end{subfigure}
    \caption{Position (left) and energy (right) of an electron starting at the origin with velocity $-1.76\cdot 10^{11}\unit{m/s}\ \textbf{e}_y$ between $t = 0\unit{s}$ and $t = 10\unit{s}$. Darker color represents larger times. Note that the energy of the electron follows an exponential pattern plus an oscillating fashion of period $\sim 5\cdot 10^{-11}\ \unit{s}$ and amplitude of the order of $10^{-18}\unit{J}$.}
    \label{fig:my_label}
\end{figure}

\section{Electromagnetic Fields}
\label{ElectromagneticFields}
\subsection{Covariant Formulation of Classical Electromagnetism}
\label{Covariant}
In Section \ref{Covariant}, we introduce the main objects and tools of the covariant formulation of classical electromagnetism, so that our work is self-contained. Throughout Section \ref{ElectromagneticFields}, Einstein's summation convention will be used unless specifically mentioned, and the considered metric in Minkowski space will be $\eta^{\alpha\mu} = \eta_{\alpha\mu} = \text{diag}(1,-1,-1,-1)$.

Recall that the four-displacement tensor is defined as $x^\mu = (ct,x,y,z)$, where $c$ is the speed of light in vacuum, and the covariant four-gradient is $\partial_\mu = \frac{\partial}{\partial^\mu} = \Big(\frac{1}{c}\frac{\partial}{\partial t}, \nabla \Big)$. If $C_\alpha$ is a four-tensor, we will use the notation $\partial_\mu C_\alpha = C_{\alpha,\mu}$.

The main object of this formulation of electromagnetism is the covariant antisymmetric tensor
\begin{equation}
	\label{EM tensor}
F^{\alpha\mu} = \begin{pmatrix}
	0 &-E_x/c& -E_y/c & -E_z/c\\
	E_x/c & 0 & -B_z & B_y\\
	E_y/c & B_z & 0 & -B_x\\
	E_z/c & -B_y & B_x & 0 
\end{pmatrix}
\end{equation}
for a pair of electric and magnetic vector fields $\textbf{E} = (E_x,E_y,E_z), \ \textbf{B} = (B_x,B_y,B_z)$. The tensor $F^{\alpha\mu}$ is known as the electromagnetic tensor. If $\phi$ and $\textbf{A}$ are a choice of scalar and vector potentials of $\textbf{E}$ and $\textbf{B}$ under the classical gauge defined in Section \ref{ClassicalGauge}, the electromagnetic four-potential is defined to be $A^\alpha = \Big(\frac{\phi}{c}, \textbf{A}\Big)$, which satisfies $F_{\alpha\mu} = \partial_\alpha A_\mu -\partial_\mu A_\alpha = A_{\mu,\alpha}-A_{\alpha,\mu}$.

Finally, an electric charge density $\rho$ and an electric current density $\textbf{j}$ define the tensor $J^\alpha = (c\rho, \textbf{j})$, which is known as the four-current. 

The four Maxwell's equations in vacuum in vector notation reduce to two tensor equations. The first, known as the Gauss-Faraday law, reads
\begin{equation}
	\label{GaussFaraday}
\partial_\mu\Big(\frac{1}{2}\epsilon^{\mu\alpha\beta\tau}F_{\beta\tau}\Big) = 0,
\end{equation}
where $\epsilon^{\mu\alpha\beta\tau}$ is the Levi-Civita tensor, and it comes from the fact that $\partial_\mu F_{\nu\lambda}+\partial_\nu F_{\lambda\mu}+\partial_\lambda F_{\mu\nu} = 0$.

The Gauss-Faraday law is the same in vacuum and in media, and thus it will not be central in our discussion. The other Maxwell's equation is known as de Gauss-Ampère law and reads
\begin{equation}
	\label{GaussAmpere}
\partial_\mu F^{\mu\alpha} = \mu_0 J^\alpha,
\end{equation}
where $\mu_0$ is the magnetic permeability of vacuum.

The Lagrangian density for classical electromagnetism is defined to be
\begin{equation}
	\label{LagrangianEM}
\mathcal{L} = -\frac{1}{4\mu_0}F^{\alpha\mu}F_{\alpha\mu}-A_\alpha J^\alpha,
\end{equation}
and it derives the Gauss-Ampère law via the Euler-Lagrange equation for fields \cite{Carroll2004}*{Ch. 1.10}. The electromagnetic energy density is defined as $u = \frac{\epsilon_0}{2}\textbf{E}\cdot\textbf{E}+\frac{1}{2\mu_0}\textbf{B}\cdot\textbf{B}$, and the energy flux density is given by the Poynting's vector field $\textbf{S} = \textbf{E}\times\frac{\textbf{B}}{\mu_0}$.

When considering electromagnetic fields in matter, the polarization density and magnetization density vector fields $\textbf{P}$ and $\textbf{M}$ encode the response of the medium to the incoming electric and magnetic vector fields, see \cite{EdwardM.Purcell2019}. The electric displacement vector is then defined as $\textbf{D} = \epsilon_0\textbf{E}+\textbf{P}$, where $\epsilon_0$ is the electric permittivity of vacuum, and the magnetic intensity is $\textbf{H} = \frac{1}{\mu_0}\textbf{B}-\textbf{M}$.

These quantities can be absorbed into the antisymmetric magnetisation-polarisation tensor
\begin{equation}
	\label{Magnetisation}
\mathcal{M}^{\alpha\mu} = \begin{pmatrix}
	0 &P_xc& P_yc & P_zc\\
	-P_xc & 0 & -M_z & M_y\\
	-P_yc & M_z & 0 & -M_x\\
	-P_zc & -M_y & M_x & 0 
\end{pmatrix}
\end{equation}
and the antisymmetric electromagnetic displacement tensor
\begin{equation}
	\label{displacement}
\mathcal{D}^{\alpha\mu} = \begin{pmatrix}
	0 &-D_xc& -D_yc & -D_zc\\
	D_xc & 0 & -H_z & H_y\\
	D_yc & H_z & 0 & -H_x\\
	D_zc & -H_y & H_x & 0 
\end{pmatrix},
\end{equation}
which are related to the electromagnetic tensor via
\begin{equation}
	\label{Relation}
\mathcal{D}^{\alpha\mu} = \frac{1}{\mu_0}F^{\alpha\mu} -\mathcal{M}^{\alpha\mu}.
\end{equation} 

Let us also recall that the bound current in a material is defined as 
\begin{equation}
	\label{boundcurrent}
J^\alpha_{bd}:=(c\rho_{bd}, \textbf{j}_{bd}): =\partial_\alpha\mathcal{M}^{\alpha\mu} = \big(-c\nabla\cdot\textbf{P}, \frac{\partial \textbf{P}}{\partial t}+\nabla\times\textbf{M}\big).
\end{equation}

As discussed above, the Gauss-Faraday law \eqref{GaussFaraday} does not change when considering fields in matter. However, the Gauss-Ampère law \eqref{GaussAmpere} becomes
\begin{equation}
	\label{GaussAmpereMatter}
	\partial_\mu\mathcal{D}^{\mu\alpha} = J^\alpha,
\end{equation}
which is known as the Gauss-Ampère law in matter. The electromagnetic energy density in a material becomes $u_{matter} = \frac{1}{2}\big(\textbf{E}\cdot\textbf{D}+\textbf{B}\cdot\textbf{H}\big)$ and the Poynting's vector is $\textbf{S}_{matter} = \textbf{E}\times\textbf{H}$.

\subsection{$k$-contact Formulation for Electromagnetic Fields}
\label{Sec: kcontactFields}
In the current section, we develop a theory that allows us to model linear and dissipative electromagnetic systems, as well as systems with contributions from both. The linear features are obtained by replacing the Minkowski metric of the spacetime with a diagonal metric that encodes information of the material, whilst the dissipative facet is modelled by introducing a linear term in the action to the Lagrangian density. Our discussion generalises the results derived in \cite{Lazo_2018}.

Recall that a linear material is such that there exist constants $\chi_e$ and $\chi_m$ for which
\begin{align}[left = \empheqbiglbrace]
	\label{linearmaterial}
	\textbf{P} &= \epsilon_0\chi_e\textbf{E}\\ 
	\textbf{M} &= \frac{1}{\mu_0}\frac{\chi_m}{1+\chi_m}: = \frac{\chi_m}{\mu}\textbf{B}\nonumber .
\end{align}

The Gauss-Ampère law for a linear material can be obtained from the electromagnetic Lagrangian density in vacuum where the Minkowski metric has been replaced. For a particular four-current $J^\alpha:U\subset\mathbb{M}^4\to Q\subset \R^4$ , define
\[
\mathcal{L} =  -\frac{1}{4\mu_0}g^{\alpha\mu}g^{\beta\nu}F_{\mu\nu}F_{\alpha\beta}-A_\alpha J^\alpha\in\mathcal{C}^\infty\Big(\bigoplus\limits_{i = 1}^4 TQ\Big),
\]
with the symmetric bilinear form 
\[
g = 	g^{\mu\nu} = \frac{1}{\sqrt{1+\chi_m}}\text{diag}\Big((1+\chi_e)(1+\chi_m),-1,-1,-1 \Big)
\]
on $U$.

The Euler-Lagrange equations for fields imply $	g^{\mu\sigma}g^{\alpha\tau}\partial_\mu F_{\sigma\tau} = \mu_0 J^\alpha$, which reads
\begin{align*}[left = \empheqbiglbrace]
    	\mu_0c\rho = \mu_0 J^0 = g^{\mu\sigma}g^{0\tau}\partial_\mu F_{\sigma\tau} = g^{00}g^{i i}\partial_i F_{i0} = \frac{1+\chi_e}{c}\nabla\cdot\textbf{E} & \iff  \nabla\cdot \textbf{D} = \rho \\
    	\mu_0 j_i = \mu_0J^i = g^{ii}g^{\mu\sigma}\partial_\mu F_{\sigma_i} = \frac{g^{ii}g^{00}}{c^2}\frac{\partial E_i}{\partial t} +(g^{ii})^2\big(\nabla\times\textbf{B}\big)_i & \iff   \textbf{j} = -\frac{\partial \textbf{D}}{\partial t}+\nabla\times\textbf{H},
\end{align*}
the non-geometric Maxwell's equations for a material satisfying \eqref{linearmaterial}.

We now add a linear term in the action to this Lagrangian density in order to model a larger subset of materials. Let
\[
\mathcal{L} =  -\frac{1}{4\mu_0}g^{\alpha\mu}g^{\beta\nu}F_{\mu\nu}F_{\alpha\beta}-A_\alpha J^\alpha-\gamma_\alpha s^\alpha\in\mathcal{C}^\infty\Big(\bigoplus\limits_{i = 1}^4 TQ\times\R\Big).
\]

Following the discussion in \cite{Gaset_2021}, the Lagrangian energy density defined by $\mathcal{L}$ is 
\[
 E_\mathcal{L} = A_{\mu,\alpha}\frac{\partial\mathcal{L}}{\partial A_{\mu,\alpha}}-\mathcal{L} = \frac{1}{\mu_0}g^{\mu\nu}g^{\alpha\beta}A_{\mu,\alpha}F_{\nu\beta}+\frac{1}{4\mu_0}g^{\mu\nu}g^{\alpha\beta}F_{\beta\nu}F_{\alpha\mu}+A_\alpha J^\alpha+\gamma_\alpha s^\alpha
\]
and gives rise to four contact $1$-forms $\theta^\alpha_{\mathcal{L}}\in\Omega\Big(\bigoplus\limits_{i = 1}^4 TQ\times \R^4\Big)$ defined by
\[
\theta_\mathcal{L}^\alpha = ds^\alpha-\frac{\partial \mathcal{L}}{\partial A_{\mu,\alpha}}dA_\mu = ds^\alpha+\frac{1}{\mu_0}g^{\alpha\beta}g^{\mu\nu}F_{\beta\nu}dA_\mu.
\]

The usual Reeb vector fields of the form $\mathcal{R}_\alpha=\frac{\partial}{\partial s^\alpha}$ fullfill the conditions
$$
\iota_{\mathcal{R}^\alpha}\theta^\beta=\delta^\beta_\alpha\,;\quad \iota_{\mathcal{R}^\alpha}d\theta^\beta=0 \,,
$$
but they are not unique. We can construct all the solutions of the previous equations by adding a general term of the form $\overline{\mathcal{R}}_\alpha=\mathcal{R}_\alpha+F_{\mu\nu,\alpha}\frac{\partial }{\partial A_{\mu,\nu}}$ with $F_{\mu\nu,\alpha}-F_{\nu\mu,\alpha}=0$. Fortunately, $\iota(\overline{\mathcal{R}}_\alpha)d E_{\mathcal{L}}=\gamma^\alpha$ for any possible (antisymetric) $F_{\mu\nu,\alpha}$, therefore the choice of Reeb vector fields does not change the equations. From now on we will use $\mathcal{R}_\alpha=\frac{\partial}{\partial s^\alpha}$.

The equations of motion for a $k$-vector field $X_\alpha$
\begin{align*}
	\iota_{X_\alpha}d\theta^\alpha_\mathcal{L} &= dE_\mathcal{L} - \gamma_\alpha\theta^\alpha_\mathcal{L}\\
	\iota_{X_\alpha}\theta^\alpha_\mathcal{L} &= - E_\mathcal{L}
\end{align*}
hence imply
\begin{align}
    \partial_\alpha s^\alpha &= \mathcal{L}\\
    F_{\tau\beta} &= A_{\beta,\tau}-A_{\tau,\beta}\\
    \mu_0 J^\mu & =g^{\nu\sigma}g^{\mu\tau}\Big(\partial_\nu F_{\sigma \tau}+\gamma_\nu F_{\sigma\tau}\Big). \label{ContactEOM}
\end{align}
Letting $\gamma^\mu = (\frac{\gamma}{c},\pmb{\gamma})$, Equation \eqref{ContactEOM} reads in vector notation as
\begin{align}
    (1+\chi_e)\big(\nabla\cdot\textbf{E}+\pmb{\gamma}\cdot\textbf{E}\big) &= \frac{\rho}{\epsilon_0}\\
    -(1+\chi_e)\epsilon_0\big(\frac{\partial\textbf{E}}{\partial t}+\gamma\textbf{E}\big)+\frac{1}{1+\chi_m}\big(\nabla\times\frac{\textbf{B}}{\mu_0}+\pmb{\gamma}\times\frac{\textbf{B}}{\mu_0}\big) &= \textbf{j}, \label{eq: maxwellcurrent}
\end{align}
which we refer to as the $k-$contact Maxwell's equations. These reduce to the Gauss-Ampère law for linear materials (or vacuum) when $\gamma_\nu = 0$.

\subsection{Generalized Poynting's Theorem}
\label{Sec: GenPoynting}

We next derive a generalized Poynting's Theorem for the obtained $k-$contact Maxwell equations in order to discuss when the modelled systems are dissipative. Dot multiplying Equation \eqref{eq: maxwellcurrent} by the vector field $\textbf{E}$ and using Faraday's law of induction ($\nabla\times\textbf{E} =-\frac{\partial \textbf{B}}{\partial t}$), we find 
\[
\textbf{E}\cdot\textbf{j} = -(1+\chi_e)\epsilon_0\Big(\frac{1}{2}\frac{\partial \textbf{E}\cdot\textbf{E}}{\partial t}+\gamma\textbf{E}\cdot\textbf{E}\Big)-\frac{1}{1+\chi_m}\Big(\frac{1}{2\mu_0}\frac{\partial\textbf{B}\cdot\textbf{B}}{\partial t}+\nabla\cdot\textbf{S}+\pmb{\gamma}\cdot\textbf{S}\Big),
\]
and integrating over a volume $V$ we obtain 
\begin{multline*}
    \int_V\frac{\partial u}{\partial t}dV+\int_{S = \partial V}\textbf{S}\cdot\hat{\textbf{n}}dS \\= -\int_V\textbf{E}\cdot\textbf{j}dV-\int_V\textbf{E}\cdot\Big(\chi_e\epsilon_0\frac{\partial \textbf{E}}{\partial t}+(1+\chi_e)\epsilon_0\gamma\textbf{E}+\frac{\chi_m}{1+\chi_m}\nabla\times\frac{\textbf{B}}{\mu_0}-\frac{1}{1+\chi_m}\pmb{\gamma}\times\frac{\textbf{B}}{\mu_0}\Big)dV,
\end{multline*}
which we refer to as the generalised Poynting's theorem. In order to shed some light on the utility of this result, let us consider the case in which $\chi_e = \chi_m = 0$. Then, the generalised Poynting's theorem reads
\[
\int_V\frac{\partial u}{\partial t}dV+\int_{S = \partial V}\textbf{S}\cdot\hat{\textbf{n}}dS = -\int_V\textbf{E}\cdot\textbf{j}dV-\int_V\textbf{E}\cdot\Big(\epsilon_0\gamma\textbf{E}-\pmb{\gamma}\times\frac{\textbf{B}}{\mu_0}\Big)dV,
\]
and the energy of the electromagnetic fields is being dissipated by both the real current $\textbf{j}$ and by a virtual current $\textbf{j}_\gamma:=\epsilon_0\gamma\textbf{E}-\pmb{\gamma}\times\frac{\textbf{B}}{\mu_0}$. Note that $\textbf{E}\cdot\textbf{j}_\gamma = \epsilon_0\gamma \textbf{E}\cdot\textbf{E}+\pmb{\gamma}\cdot\textbf{S}$. Whenever the vector $\pmb{\gamma}$ vanishes and $\gamma>0$, then the volume integral
\[
\int_V\textbf{E}\cdot\textbf{j}_\gamma dV = \epsilon_0\gamma\int_V\textbf{E}\cdot\textbf{E}dV
\]
is positive, and thus the modelled system is indeed dissipative. This discussion agrees with that made in \cite{Lazo_2018}. In general, the virtual current is $\textbf{j}_\gamma = \chi_e\epsilon_0\frac{\partial \textbf{E}}{\partial t}+(1+\chi_e)\epsilon_0\gamma\textbf{E}+\frac{\chi_m}{1+\chi_m}\nabla\times\frac{\textbf{B}}{\mu_0}-\frac{1}{1+\chi_m}\pmb{\gamma}\times\frac{\textbf{B}}{\mu_0}$, and
\[
\textbf{E}\cdot\textbf{j}_\gamma = \frac{1}{2}\frac{\partial}{\partial t}\Big(\chi_e\epsilon_0\textbf{E}\cdot\textbf{E}-\frac{\chi_m}{1+\chi_m}\frac{1}{\mu_0}\textbf{B}\cdot\textbf{B}\Big)+(1+\chi_e)\textbf{E}\cdot\textbf{E}+\frac{\chi_m}{1+\chi_m}\nabla\cdot\textbf{S}+\frac{1}{1+\chi_m}\pmb{\gamma}\cdot\textbf{S}.
\]

One can also argue as follows. Following the constitutive relations for $\textbf{D}$ and $\textbf{H}$ in a linear material, let $\tilde{\textbf{D}}:=(1+\chi_e)\epsilon_0\textbf{E}$ and $\tilde{\textbf{H}} := \frac{1}{1+\chi_m}\frac{1}{\mu_0}\textbf{B}$. Then, define $\tilde{u} :=\frac{1}{2}\Big(\textbf{E}\cdot\tilde{\textbf{D}}+\textbf{B}\cdot\tilde{\textbf{H}}\Big) = \frac{1}{2}\Big((1+\chi_e)\epsilon_0 \textbf{E}\cdot \textbf{E}+\frac{1}{1+\chi_m}\frac{1}{\mu_0}\textbf{B}\cdot\textbf{B}\Big)$ and $\tilde{\textbf{S}}: = \textbf{E}\times\tilde{\textbf{H}} = \frac{1}{1+\chi_m}\textbf{S}$. Then, the generalized Poynting's theorem implies
\begin{multline*}
    \int_V\frac{\partial \tilde{u}}{\partial t}dV+\int_{S = \partial V}\tilde{\textbf{S}}\cdot\hat{\textbf{n}}dS = -\int_V\textbf{E}\cdot\textbf{j}dV-\int_V\textbf{E}\cdot\Big(\gamma\tilde{\textbf{D}}-\pmb{\gamma}\times\tilde{\textbf{H}}\Big)dV,
\end{multline*}
and the virtual current dissipating energy from the system becomes $\tilde{\textbf{j}}_\gamma = \gamma\tilde{\textbf{D}}-\pmb{\gamma}\times\tilde{\textbf{H}}$. 

\subsection{$k$-contact Maxwell's equations as Maxwell's equations in matter}

As discussed at the end of Section \ref{Sec: kcontactFields}, the $k$-contact Maxwell's equations are precisely Maxwell's equations for a linear material or vacuum when $\gamma_\nu = 0$. In general, imposing that the obtained equations of motion are precisely the Gauss-Ampère law is equivalent to the existence of constants $\chi_e$ and $\chi_m$ and a four tensor $\gamma^\nu$ for which the identity $g^{\nu\sigma}g^{\mu\tau}\Big(\partial_\nu F_{\sigma\tau}+\gamma\nu F_{\sigma\tau}\Big) = \mu_0\eta^{\nu\alpha}\eta^{\mu\beta}\partial_\nu\mathcal{D}_{\alpha\beta}$ is satisfied. Looking at the $k$-contact Maxwell's equations, it is sufficient that the material satisfies 
\begin{align}
    \nabla\cdot\textbf{P} &=\chi_e \epsilon_0 \nabla\cdot\nonumber \textbf{E}+(1+\chi_e)\epsilon_0 \pmb{\gamma}\cdot\textbf{E}\\\label{Eq: conditions}
     \frac{\partial\textbf{P}}{\partial t} &= \chi_e\epsilon_0\frac{\partial\textbf{E}}{\partial t}+(1+\chi_e)\epsilon_0\gamma\textbf{E}\\
    \nabla\times \textbf{M} & =\frac{\chi_m}{1+\chi_m}\nabla\times\frac{\textbf{B}}{\mu_0}\nonumber -\frac{1}{1+\chi_m}\pmb{\gamma}\times\frac{\textbf{B}}{\mu_0}
\end{align}
for certain constants $\chi_e, \chi_m,\gamma\in\R$ and $\pmb{\gamma}\in\R^3$. Note that it is necessary that the incoming electric field fulfils $\pmb{\gamma}\cdot\frac{\partial\textbf{E}}{\partial t} = \gamma_0\nabla\cdot\textbf{E}$.

Note that whenever these relations are satisfied, and hence our framework models electromagnetism in matter, the bound current $\textbf{j}_{bd}$ induced by the material is precisely the virtual current appearing in the generalized Poynting's theorem in Section \ref{Sec: GenPoynting}, that is
\[
\textbf{j}_{bd} = \frac{\partial \textbf{P}}{\partial t}+\nabla\times\textbf{M} = \chi_e\epsilon_0\frac{\partial\textbf{E}}{\partial t}+(1+\chi_e)\epsilon_0\gamma\textbf{E} + \frac{\chi_m}{1+\chi_m}\nabla\times\frac{\textbf{B}}{\mu_0} -\frac{1}{1+\chi_m}\pmb{\gamma}\times\frac{\textbf{B}}{\mu_0} = \textbf{j}_\gamma.
\]

Hence, our framework models dissipative electromagnetic systems whenever the bound current $\textbf{j}_{bd} = \textbf{j}_\gamma$ is a dissipative term, as seen in the generalized Poynting's Theorem.

\subsection{Example. The Lorentz Dipole Model}

The Lorentz dipole models the interaction between an oscillating electric field and an electron \cite{Hecht2017}*{Ch. 3}. The model assumes the electron behaves like a harmonic oscillator attached to the nucleus by a hypothetical spring of constant $C$, and that the oscillations are driven by the electric field. The source of damping is not specified but comes from a drag force $\vec{F} = -\mu\vec{v}_e$ on the electron. Let $m$ be the mass of the electron and $q_e>0$ the absolute value of it's charge.

The solution to the equations of motion can be found assuming the incoming field to be of the form $\textbf{E}(\textbf{r}, t) = \textbf{E}_0e^{i\textbf{k}\cdot\textbf{r}-i\omega t}$, where $\textbf{k}$ is called the wave vector and $\omega$ is the frequency. One can derive the constitutive relation of the polarization $\textbf{P} = \epsilon_0(\chi'-i\chi'')\textbf{E}$ of a material formed of $N$ such atoms that do not interact with each other, where the electric susceptibility becomes a complex quantity with
\[
\chi' = \omega_p^2\frac{(\omega_0^2-\omega^2)}{(\omega_0^2-\omega^2)^2+(\omega/\tau)^2}\hspace*{3cm} \chi'' = \omega_p^2\frac{\omega/\tau}{(\omega_0^2-\omega^2)^2+(\omega/\tau)^2},
\]
where $\omega_0 = \sqrt{\frac{C}{m}}$ is the natural frequency of the electron-nucleus system and $\omega_p^2 = \frac{Nq_e}{\epsilon_0 m}$. We have also defined $\tau = \frac{m}{\mu}$. The complex part of the susceptibility models the absorption of light by the material, and hence the dissipation of electromagnetic energy of the incoming field.

Note that 
\begin{align*}
\frac{\partial \textbf{P}}{\partial t} = \epsilon_0(\chi'-i\chi'')\frac{\partial \textbf{E}}{\partial t} =& \epsilon_0\chi'\frac{\partial\textbf{E}}{\partial t}-\epsilon_0\chi''\omega\textbf{E}\\
\nabla\cdot\textbf{P} = \epsilon_0(\chi'-i\chi'')\nabla\cdot\textbf{E} &= \epsilon_0\chi'\nabla\cdot\textbf{E}+\epsilon_0\chi''\textbf{k}\cdot\textbf{E},
\end{align*}
and since no magnetic effects are considered, it is sufficient to define $\chi_e = \chi'$, $\chi_m = 0$ and $\gamma = -\frac{\chi''}{1+\chi'}\omega$, $\pmb{\gamma} = \frac{\chi''}{1+\chi'}\textbf{k}$ for Equations \eqref{Eq: conditions} to be satisfied.

\subsection{Example. Highly Resistive Dielectric}
Consider a parallel capacitor filled with a dielectric. The induced vector field is $\textbf{E} = \frac{\Delta V}{d}\textbf{e}_z$, where $\Delta V$ is the constant applied voltage difference, $d$ is the distance between the plates and $\textbf{e}_z$ is the unit vector orthogonal to the plates. Assume no external electric or magnetic fields intervene and that the dielectric is non-magnetic.

Assume that the dielectric is composed of spherical molecules with a high resistivity. These are of radius $R$ and separated a distance $s\gg R$. The electric field inside the dielectric away from the centre of the spheres is essentially constant and equal to $\frac{\Delta V}{d}e_\textbf{z}$.

Due to the high resistivity of the molecules, the accumulation of charge at the poles as a response to the applied electric field is described as a rate, rather than a magnitude, which is proportional to the external electric field, see \cite{Haus1989}*{Sec. 11.5}. Thus, we expect $\frac{\partial\textbf{P}}{\partial t}$ to be proportional to $\textbf{E}$, contrary to when the spheres are perfect conductors, which makes the dielectric linear and hence $\textbf{P}$ proportional to $\textbf{E}$. Thus, the polarization constitutive law for a short initial period of time reads
\[
\frac{\partial\textbf{P}}{\partial t} = \alpha\textbf{E},
\]
where $\alpha$ is a positive real constant dependant only on the properties and geometry of the spheres.

Hence, one can take $\chi_e = \chi_m = 0$ and $\gamma = \frac{\alpha}{\epsilon_0 c}$. Also, if no interaction between the spheres is considered, $\nabla\cdot \textbf{D} = 0$ and $\nabla\cdot \textbf{P}= 0$. Also, since the magnetization of the molecules is zero, $\nabla\times\textbf{M}= 0$ and if we let $\pmb{\gamma} = 0$,
\[0 = \pmb{\gamma}\cdot\textbf{E} = \nabla\cdot\textbf{P} = 0\]
 and 
\[0=-\pmb{\gamma}\times\textbf{B} =\mu_0\nabla\times\textbf{M} = 0\]
and hence Equations \eqref{Eq: conditions} are satisfied. Since $\pmb{\gamma} = 0$ and $\gamma_0>0$, the bound current is a source of dissipation, as discussed in Section \ref{Sec: GenPoynting}. This coincides with the conclusions drawn in \cite{Haus1989}*{Sec. 11.5} by means of physical arguments.

\subsection{Gauge Invariance}
\label{GaugeInvariance}
In the current section we show how the developed $k$-contact theory for electromagnetic fields is invariant under the classical gauge theory for fields. 

Let $Q = \R^4$ be the space in which the four tensor $A_\mu$ takes its values, and consider a general function $f:Q\rightarrow \R$. The change of gauge given by $f$ is 
\[
A'_\mu = A_\mu-\partial_\mu f.
\]
Let us consider the case in which the external current $J^\alpha$ vanishes. Let the primed variables denote the objects in the new gauge, and the unprimed variables denote the objects in the old gauge. Note first that
\[
F'_{\alpha\mu} = \partial_\alpha A'_\mu-\partial_\mu A'_\alpha = F_{\alpha\mu} -\partial_{\alpha\mu}f+\partial_{\alpha\mu}f = F_{\alpha\mu},
\]
and hence $F_{\alpha\mu}$ is invariant under the gauge. Thus, the Lagrangian
\[
\mathcal{L} = -\frac{1}{4\mu_0}g^{\alpha\mu}g^{\beta\nu}F_{\mu\nu}F_{\alpha\beta}-\gamma_\alpha s^\alpha
\]
is invariant under the gauge. Also, the energy satisfies
\[
E'_\mathcal{L}  = \frac{1}{\mu_0}g^{\mu\nu}g^{\alpha\beta}A'_{\mu,\alpha}F'_{\nu\beta}+\frac{1}{4\mu_0}g^{\mu\nu}g^{\alpha\beta}F'_{\beta\nu}F'_{\alpha\mu}+\gamma_\alpha s^\alpha = E_\mathcal{L}-\frac{1}{\mu_0}g^{\mu\nu}g^{\alpha\beta} F_{\nu\beta}\partial_{\alpha\mu}f
\]
where the last equality follows from the fact that the tensor $\partial_{\alpha\mu}f$ is symmetric while $F^{\alpha\mu}$ is antisymmetric.

It is also clear that the equations of motion
\[
0=g^{\nu\sigma}g^{\mu\tau}\Big(\partial_\nu F_{\sigma \tau}+\gamma_\nu F_{\sigma\tau}\Big)
\]
are invariant under the gauge.

Thus, our $k$-contact theory and all of the objects involved are invariant under the classical gauge theory for electromagnetic fields.

\section{Conclusions}

The work presented here contributes to the understanding of how contact geometry can be applied in classical mechanics and classical field theory to describe systems with dissipation and damping. A lot of research has been done lately to build a general theory, discovering conservation and dissipation theorems and analogues to classical results such as Noether's theorem \cites{de_Le_n_2019singularandprecontact, Gaset_2020contactframework, Gaset_2021, Gaset_2020newcon}, but no extensive studies on applications to current theories and real-life examples have been made apart from small examples in \cites{Gaset_2020newcon, Lazo_2018, de_Le_n_2019singularandprecontact, Gaset_2020contactframework, Gaset_2021}.

The present work focuses on how contact formalism can be applied to electromagnetic systems with dissipation. It has been studied how the techniques introduced by G. Herglotz in 1930, and recently formalized using the language of contact geometry, can be applied to particles under the influence of a Lorentz force and external damping. Moreover, a recent theory developed in \cites{Gaset_2020contactframework, Gaset_2021}, which generalizes contact formalism to study classical field theory, has been used to discuss electromagnetic fields themselves when being dissipated by external phenomena.

 The contact formalism of classical mechanics deals with systems whose Lagrangian function or density depends on the action itself. Driven by previous successful applications of the theory, in the present work the Lagrangians under study have been taken to be the classical Lagrangians of electromagnetism (the Lagrangian of the Lorentz force and the Lagrangian density of electromagnetic fields) plus a linear term in the action.
 
 When applying the theory to particles under the Lorentz force, the produced equations of motion have turned out to be not invariant under the current gauge theory. This has driven us to propose a new gauge approach for electromagnetism which generalizes the classical one and reduces to the currently accepted theory for non-dissipative symplectic systems. It has also been necessary to slightly change the Lagrangian of the Lorentz force, adding an extra term which produces the same equations of motion in the non-dissipative case but that is key in the contact formalism. All of these generalizations make a new mathematical observable vector field appear, which can be seen to vanish from the equations of motion in the symplectic case. Moreover, in this new paradigm, the generalized momentum of a particle under a Lorentz force becomes an observable, whilst in the symplectic theory of electromagnetism the generalized momenta are not gauge-invariant. 

The developed approach for particles under a Lorentz force in the presence of external damping has been applied to two real-life scenarios. Firstly, our theory is able to model the behaviour of an electron inside a non-perfect conductor. When imposing that the limit velocity of the particle agrees with the current theory, Joules' heating law is recovered. It has also been discussed how our description can be applied to a particle in a magnetic field with dissipation.

When considering vector fields, we have produced a new set of Gauss-Ampère equations which reduce to the classical Gauss-Ampère equation when omitting the dissipation parameters. In the general case, we have showed how our theory models electromagnetic fields in matter for a certain type of systems. It has been argued how these systems might have in common that the bound four-current generated by the material is precisely a dissipative term. Further research is needed in order to better understand such systems and characterize them further. Our theory has been seen to be able to describe a particular regime of the Lorentz dipole model.

This work pretends to be a first approach towards the use of contact geometry techniques to model dissipation in electromagnetism. It should be further studied whether the generalizations needed to make the theory gauge invariant for the Lorentz force can actually be seen as generalizations of the current theory of electromagnetism. For fields, more research is needed to better understand how the modelled systems behave and find whether other variations of the Lagrangian can describe different systems. It is also necessary to produce more and more relevant examples of how our equations can be applied to real problems in Physics and reproduce current data and predict new phenomena. Furthermore, the contact framework needs to be further developed so that it admits non-autonomous Lagrangians which depend explicitly on time or the components of spacetime, as these arise naturally in many theories in Physics.

\bibliographystyle{abbrv}
\bibliography{ContactPaper}

\end{document}